\documentstyle[preprint,aps]{revtex}
\begin{document}
\draft
\preprint{IFUSP-P/1183, hep-th/9510136}
\title{THE SEVERAL GUISES OF THE BRST SYMMETRY}
\author{Victor O. Rivelles\cite{email}}
\address{Instituto de F\'{\i}sica, Universidade de S\~ao Paulo,\\
Caixa Postal 66318, 05389-970 S\~ao Paulo, SP, Brazil}
\date{October 1995}
\maketitle

\begin{abstract}
We present several forms in which the BRST transformations of QCD in
covariant gauges can be cast. They can be non-local and even not
manifestly covariant. These transformations may be obtained in the
path integral formalism by non standard integrations in the ghost
sector or
by performing
changes of ghost variables which leave the action and the path integral
measure invariant. For different changes of ghost variables in the BRST
and anti-BRST transformations these two transformations  no longer
anticommute.
\end{abstract}

\section{Introduction}
\label{sec:intro}
BRST symmetry \cite{brst} is by now a familiar concept associated to
to the quantization of any gauge theory. It reflects in an deep
way the gauge symmetry at the quantum level. For QCD in a covariant
gauge there is a standard form for the BRST transformations
\begin{eqnarray}
\label{brst-standard}
\delta A^a_\mu &=& D_\mu c^a \nonumber \\
\delta c^a &=& - \frac{1}{2} g f^{abc} c^b c^c \nonumber \\
\delta \overline{c}^a &=& - \frac{i}{\xi} \partial_\mu A^{\mu a} \nonumber \\
\delta \psi &=& - g  c^a \lambda^a \psi
\end{eqnarray}
This form however is not unique. In fact several new symmetries of the
quantum action of QED in Feynman gauge have been reported
\cite{lm,tang,yang}. It has
been pointed out however that they are just BRST transformations in
non standard form \cite{mine}. In this paper we will elaborate on this
point to extended it to the non-abelian case. We will show how the
search for non standard BRST transformations, and by this we mean non
local and/or not manifestly covariant BRST transformations, can be
performed in a systematic way.

A reason for the existence of different forms of the BRST
transformations is the fact that there is a set of changes of
variables, not
necessarily local, that leaves the ghost action
invariant. There is also some freedom in the canonical
formalism. For example, in the Batalin-Fradkin-Vilkovisky (BFV) formalism
\cite{bfv} we usually perform the path integration over the ghosts momenta
to arrive at the transformations Eqs.(\ref{brst-standard}). We could instead
perform the integration over the ghosts themselves and leave the ghost
momenta in the  action. The resulting action is local after some
changes of variables but the BRST transformations are in general
non-local and not manifestly covariant.

In Section \ref{sec:1} we make a brief presentation of the BFV
formalism applied to QCD to set up our conventions and set the stage
for the next section. Then in Section \ref{sec:2} we derive the BRST
transformations when we perform the alternative ghost path integration
mentioned above. We show that there are two sets of BRST transformations,
the usual one which is covariant and local Eqs.(\ref{brst-standard})
and another one which is not manifestly covariant and non local.
In the next section we consider changes of ghost variables
which leave the action and the path integral measure invariant. We
particularize to the abelian case to avoid the unnecessary algebraic
complications of the non-abelian structure. We show in particular that
there exist changes of ghost variables which lead to BRST and
anti-BRST transformations which do not anticommute. In the BFV
formalism these changes of variables correspond to canonical
transformations in the ghost sector. At last in Section
\ref{sec:end} we make some final comments.

\section{Resum\'e of the BFV Formalism for QCD}
\label{sec:1}

We will present in a very summarized way how the usual BRST transformations
are obtained in BFV formalism \cite{bfv}.

The first step is to find out the constraint structure of QCD. After
that we introduce the ghosts and their momenta to build up the BRST
charge $Q$.
The quantum QCD action is then\cite{conv}
\begin{equation}
\label{action0}
S_q = \int d^4 x \,\,\, ( \Pi^{ia} \dot A^a_i + \Pi^{0a} \dot A^a_0 +
i \overline \psi \gamma^0 \dot \psi + \dot{\cal P}^a \overline{c}^a +
\dot{c}^a \overline{\cal P}^a - H_c - \{ \Psi, Q \} )
\end{equation}
where $\Pi^{\mu a}$ are the momenta conjugated to $A^a_\mu$;  ${\cal
P}^a, \overline{\cal P}^a$ are the ghost momenta conjugated to the
ghosts $\overline{c}^a, c^a$; $H_c$ is the canonical QCD Hamiltonian and
$\Psi$ is the gauge fixing fermion. The BRST charge is given by
\begin{equation}
\label{Q}
Q = \int d^3 x \,\,\, [ ( D_i F^{i0 a} + i g J^a_0 ) c^a + \frac{1}{2}
f^{abc} \overline{\cal P}^a c^b c^c - i {\cal P}^a \Pi^{0 a} ]
\end{equation}
The main assertion of the BFV formalism is that the
path integral $Z = \int D[\phi] \,\,\, \exp i S_q $ is independent of the gauge
fixing fermion. Here $D[ \phi ]$ is the usual Liouville measure over all
fields and ghosts.

Proper choices of $\Psi$ allows to recover the
usual gauge conditions. Covariant gauges are implemented by the
choice
\begin{equation}
\label{psi}
\Psi = \int d^3 x \,\,\, [ i \overline{c}^a (
\frac{1}{2} \xi \Pi^{0a} + \partial_i A^{ia} ) + \overline{\cal P}^a
A^a_0 ]
\end{equation}
For $\xi = 1$ we get the
Feynman gauge, $\xi = 0$ the Landau gauge and $\xi \rightarrow
\infty$ the unitary gauge.  By performing the
functional integration over the momenta $\Pi^{\mu a}$ we arrive at the
effective action $S$
\begin{eqnarray}
\label{action1}
S &=& S_{QCD} + S_{gf} + S_{gh} \nonumber \\
S_{QCD}  &=& \int d^4x \,\,\, [ - \frac{1}{4} F^a_{\mu \nu} F^{\mu \nu a} +
\overline{\psi} ( i \gamma^\mu D_\mu - m ) \psi ] \nonumber \\
S_{gf} &=& - \int d^4x \,\,\, \frac{1}{2 \xi} ( \partial_\mu A^{\mu a})
( \partial_\nu A^{\nu a} ) \nonumber \\
S_{gh} &=& \int d^4x \,\,\, ( i \overline{c}^a \partial_i D^i c^a +
\dot{\cal P}^a \overline{c}^a - \overline{\cal P}^a D_0 c^a +
i \overline{\cal P}^a {\cal P}^a )
\end{eqnarray}
where $S_{QCD}$ is the classical QCD action, $S_{gf}$ is the gauge
fixing action and $S_{gh}$ is the ghost action. The effective action
$S$ is then
invariant under the BRST
transformations generated
by $Q$ which have the form
\begin{eqnarray}
\label{brst1}
&& \delta A^a_i = D_i c^a, \,\,\,\,\,\, \delta A^a_0 = i {\cal P}^a
\nonumber \\
&& \delta c^a = - \frac{1}{2} g f^{abc} c^b c^c, \,\,\,\,\,\,  \delta
\overline{c}^a = - \frac{i}{\xi} \partial_\mu A^{\mu a} \nonumber \\
&& \delta {\cal P}^a = 0, \,\,\,\,\,\, \delta \overline{\cal P}^a =
D_i F^{0i a} -
i g J^a_0 - g f^{abc} \overline{\cal P}^b c^c \nonumber \\
&& \delta \psi = - g c^a \lambda^a \psi
\end{eqnarray}
where $J^a_0 = \overline{\psi} \lambda^a \gamma_0 \psi$ is the time
component of the current.

At this stage we usually perform the path integration over ${\cal P}^a$ and
$\overline{\cal P}^a$. Integrating over ${\cal P}^a$ gives a delta
functional $\delta( i \overline{\cal P}^a + \dot{\overline{c}}^a)$
and integrating over $\overline{\cal P}^a$ allow us to replace
$\overline{\cal P}^a$ by $i \dot{\overline{c}}^a$.
Then $S_{gh}$ in Eqs.(\ref{action1}) gets its usual Faddeev-Popov form
\begin{equation}
\label{action-gh1}
S_{gh} = \int d^4x \,\,\,i \overline{c}^a \partial^\mu D_\mu c^a
\end{equation}
and we get the usual BRST transformations
Eqs.(\ref{brst-standard}). Notice
the importance
of this last two integrals. We restore manifest covariance in
the BRST transformations Eqs.(\ref{brst1})
and we get a local and covariant ghost action
Eq.(\ref{action-gh1}). In the next section we
will describe the
results when we perform the integrations on other pair of ghost variables.

As usual the BRST transformations are nilpotent
on-shell. The nilpotency fails only on
$\overline{c}^a$ and is proportional to the $c^a$ field
equation. Usually this can be overcome by the introduction of an
auxiliary field.  In the BFV formalism, however, it corresponds to the
situation where it is not performed the path integration over
$\Pi^a_0$. If we call $\lambda^a = \Pi^a_0 + \frac{1}{\xi}
\partial_\mu A^{\mu a}$ then the gauge fixed action has one more term
and becomes
\begin{equation}
\label{action-aux}
S_{gf} = \int d^4 x \,\,\, [ - \frac{1}{2 \xi} (\partial_\mu A^{\mu
a})^2 + \frac{1}{2} \xi \lambda^a \lambda^a ]
\end{equation}
and the BRST transformations are the same as Eqs.(\ref{brst-standard})
before except for
$\overline{c}^a$. The new BRST transformations are then
\begin{eqnarray}
\label{brst-aux}
&& \delta \overline{c}^a = - \frac{i}{\xi} \partial_\mu A^{\mu a} + i
\lambda^a \nonumber \\
&& \delta \lambda^a = \frac{1}{\xi} \partial^\mu D_\mu c^a
\end{eqnarray}
In Section \ref{sec:3} we will consider the case of QED with this
extra field.

Besides the BRST transformations the quantum action is also invariant under
anti-BRST transformations. In the BFV formalism the anti-BRST charge
can be obtained from the BRST charge by interchanging ghosts by
anti-ghosts in such a way that both charges anticommute. We then find
that the anti-BRST transformations in a covariant gauge are
\begin{eqnarray}
\label{antibrst1}
\overline{\delta} A^a_\mu &=& D_\mu \overline{c}^a \nonumber \\
\overline{\delta} c^a &=& - \frac{i}{\xi} \partial_\mu A^{\mu a} \nonumber \\
\overline{\delta} \overline{c}^a &=& \frac{1}{2} g f^{abc}
\overline{c}^b \overline{c}^c \nonumber \\
\overline{\delta} \psi &=& - g \overline{c}^a \lambda^a \psi
\end{eqnarray}
It should be noticed that the anti-BRST symmetry is not a new
symmetry. The anti-BRST charge has the same information content as the
BRST charge and we could use anyone to generate physical states or
to obtain Ward identities. Often both are used.

\section{Non-local BRST transformations}
\label{sec:2}

Consider the BFV formalism in the previous section up to the
point where the integration over $\Pi^{\mu a}$ was performed and
Eqs.(\ref{action1},\ref{brst1}) were obtained.
We will now perform the integration over the ghost fields instead of
their momenta. Integrating over $\overline{c}^a$ gives $\delta( i
\partial_i D^i c^a - \dot{\cal P}^a) = det( \partial_i D^i ) \,\,\,
\delta( i c^a - \frac{1}{\partial_i D^i} \dot{\cal P}^a)$. Integrating
now  over $c^a$ we can replace $c^a = - i \frac{1}{\partial_i D^i}
\dot{\cal P}^a$. Then the ghost action in Eqs.(\ref{action1}) becomes
\begin{equation}
\label{action2}
S_{gh} = \int d^4x \,\,\, ( i \overline{\cal P}^a D_0
\frac{1}{\partial_i D^i} \dot{\cal P}^a + i \overline{\cal P}^a {\cal
P}^a )
\end{equation}
Notice the appearance of the non-local term in the ghost action as a result
of this unusual integration. Notice also that the path integral
measure has now an extra term $det( \partial_i D^i )$ which should be
taken into account. We can overcome these two troubles by making
judicious changes of variables. First perform the change of variables
$\dot{\cal P}^a \rightarrow i c^a, \overline{\cal P}^a \rightarrow
i \dot{\overline{c}}^a$ whose Jacobian is one. Then perform a second
change of variables $c^a \rightarrow - \partial_i D^i c^a$ whose
Jacobian is $det^{-1}(\partial_i D^i)$. As a result the contribution
from the Jacobian to the path integral measure
cancels out the contribution from the ghost integration. Also the
non-local ghost
action Eq.(\ref{action2}) becomes local and it takes the usual
Faddeev-Popov form
Eq.(\ref{action-gh1}). The BRST transformations can now be obtained from
Eqs.(\ref{brst1}). They are
\begin{eqnarray}
\label{brst3}
&& \delta A^a_i = D_i c^a, \,\,\,\,\,\, \delta A^a_0 = - \frac{1}{\partial_0}
\partial_j D^j c^a \nonumber \\
&& \delta c^a = - \frac{1}{2} g f^{abc} c^b c^c \nonumber \\
&& \delta \overline{c}^a =  - i \frac{1}{\partial_0} D_i F^{0i a}
- g f^{abc} \frac{1}{\partial_0} ( \dot{\overline{c}^b} c^c ) - g
\frac{1}{\partial_0} J^a_0 \nonumber \\
&& \delta \psi = - g c^a \lambda^a \psi
\end{eqnarray}
We end up then with a local ghost action and a set of non-local and
not manifestly covariant BRST transformations by performing the integration
over
the ghost fields instead of their momenta. The transformations
Eqs.(\ref{brst3}) leave the effective action invariant and are nilpotent
as any good BRST transformation.

There are two other possibilities to perform the ghost integrations.
If we perform the path integration over $\overline{c}^a$ and
${\cal P}^a$ we obtain the same result as before after proper changes
of variables. If we integrate over $\overline{\cal P}^a$ and $ c^a$
instead
we get the usual BRST transformations Eqs.(\ref{brst-standard})
also after proper change of variables.

Then in the BFV formalism we can arrive at a local ghost action, the
Faddeev-Popov action,  and
two standard sets of BRST
transformations which can be either covariant and local
Eqs.(\ref{brst-standard}) or not manifestly covariant and non-local
Eqs.(\ref{brst3}). It should also be noticed that both sets of BRST
transformations reduce to each other on-shell. If we use the $c^a$
and $A^a_0$ field equations we can turn the non-local BRST
transformations Eqs.(\ref{brst3}) into the local ones
Eqs.(\ref{brst-standard}).


\section{Changes of Variables in the Ghost Action}
\label{sec:3}

{}From now on let us consider just the abelian case for the sake of
simplicity. The ghost action is then $S_{gh} = \int d^4x \,\,\, i
\overline{c} \Box c$. This action allows a huge freedom to
perform changes of variables which leave the path integration measure
and the action itself
invariant.

Let us consider first some cases with the non-local form of the BRST
transformations.
If we perform the following change of variables $c \rightarrow
i \frac{\partial_0}{ \nabla^2 } c $ and $  \overline{c} \rightarrow i
\frac{\nabla^2 }{\partial_0} \overline{c}$ whose Jacobian is one,
then Eqs.(\ref{brst3}) in the abelian case reduce to
\begin{eqnarray}
\label{brst-lm}
&&\delta A_i = i \frac{\partial_i}{\nabla^2} \dot{ c}, \,\,\,\,\,\,
\delta A_0 = i c \nonumber \\
&&\delta c = 0, \,\,\,\,\,\,  \delta \overline{c} =  \frac{1}{\nabla^2}
\partial_i \dot{A}_i - A_0 + i g \frac{1}{\nabla^2} J_0 \nonumber \\
&&\delta \psi = - i g (\frac{1}{\nabla^2}\dot{ c}) \psi
\end{eqnarray}
These are precisely the transformations found in Ref.\cite{lm}. It
has also been pointed out that they can be found by a canonical
transformation in the ghost sector before any integration is performed
\cite{gaete}.

Another change of variables $c \rightarrow -  \dot{c} $ and $
\dot{\overline{c}} \rightarrow -  \overline{c}$ (which also has
Jacobian one) reduces the abelian form of Eqs.(\ref{brst3}) to
\begin{eqnarray}
\label{brst-yang}
&& \delta A_i = - \partial_i \dot{c}, \,\,\,\,\, \delta A_0 = -
\nabla^2 c \nonumber \\
&&\delta c = 0, \,\,\,\,\,\, \delta \overline{c} = i \partial_i F^{0i}
+ g J_0 \nonumber \\
&&\delta \psi = g \dot{c} \psi
\end{eqnarray}
We then get the local but not manifestly covariant transformations found in
Ref.\cite{yang}.

Still another change of variables $c \rightarrow
\frac{\partial_0}{\nabla^2} c $ and $ \overline{c} \rightarrow
\frac{\nabla^2}{\partial_0} \overline{c}$ this time in the abelian
form of the usual BRST
transformations Eqs.(\ref{brst-standard}) produces
\begin{eqnarray}
\label{brst-fink}
&& \delta A_\mu = D_\mu \frac{1}{\nabla^2} \dot{c}  \nonumber \\
&& \delta c = 0, \,\,\, \delta \overline{c} = -\frac{i}{\xi}
\frac{1}{\nabla^2} \partial_\mu \dot{A}^\mu  \nonumber \\
&& \delta \psi = - g ( \frac{1}{\nabla^2} \dot{c} ) \psi
\end{eqnarray}
These are the non-local transformations found in
Ref.\cite{tang}.

This procedure of performing changes of variables on the ghost fields
can be easily generalized. Assume that $F, G, \dots$ are operators
which possess an adjoint $F^t, G^t, \dots$ in the sense that
\begin{equation}
\label{adj}
\int dx  \,\,\, \phi \,\, F[ \psi ] = \int  dx \,\,\, F^t[ \phi ] \,\,
\psi
\end{equation}
Notice that the effective action in the abelian case defines a bilinear
metric $\int dx \,\, \phi \,\, \psi$. Assume also that these operators
are field independent so that they
commute with $\partial_\mu$.
Let us consider also the extra field which makes the BRST
transformations nilpotent off-shell as mentioned at the end of
Section \ref{sec:1}. Consider now the change of variables
\begin{eqnarray}
\label{F}
&& A_\mu \rightarrow A_\mu, \,\,\,\,\,\,  \lambda \rightarrow \lambda,
\,\,\,\,\,\, \psi \rightarrow \psi \nonumber \\
&& c \rightarrow F[c],\,\,\,  \overline{c} \rightarrow (F^{-1})^t [
\overline{c} ]
\end{eqnarray}
whose Jacobian
is one. Let us call this change of variables a $F$
transformation. Then the abelian ghost action remains
invariant under this $F$ transformation and the local abelian BRST
transformations Eqs.(\ref{brst-standard}, \ref{brst-aux})
become
\begin{eqnarray}
\label{brst-f}
&& \delta A_\mu = \partial_\mu F[c], \,\,\,\,\,\, \delta \lambda =
\frac{1}{\xi} \Box F[c] \nonumber \\
&& \delta c = 0, \,\,\,\,\,\, \delta \overline{c} = F^t[ - \frac{i}{\xi}
\partial^\mu A_\mu + i \lambda ] \nonumber \\
&& \delta \psi = g \psi F[c]
\end{eqnarray}
and clearly generalizes Eqs.(\ref{brst-fink}).
If instead we perform the change of variables in the
abelian non-local BRST transformations Eqs.(\ref{brst3}) we get the
generalization of the BRST transformations Eqs.(\ref{brst-lm},
\ref{brst-yang}).

We can now consider the following
situation. Consider a $F$ transformation and the resulting BRST
transformations Eqs.(\ref{brst-f}). Consider a $G$ transformation
defined as
\begin{eqnarray}
\label{G}
&&  A_\mu \rightarrow A_\mu, \,\,\,\,\,\,  \lambda \rightarrow
\lambda,
\,\,\,\,\,\, \psi \rightarrow \psi \nonumber \\
&& c \rightarrow ( G^{-1} )^t[ c ], \,\,\,\,\, \overline{c} \rightarrow
G [ \overline{c} ]
\end{eqnarray}
and the resulting {\bf anti}-BRST transformations, that is,
\begin{eqnarray}
\label{brst-g}
&& \overline{\delta} A_\mu = \partial_\mu G [ c ], \,\,\,\,\,\,
\overline{\delta} \lambda = \frac{1}{\xi} \Box G [ \overline{c} ]
\nonumber \\
&& \overline{\delta} c = G [  - \frac{i}{\xi}
\partial^\mu A_\mu + i \lambda ] \,\,\,\,\,\, \overline{\delta}
\overline{c} = 0 \nonumber \\
&& \overline{\delta} \psi = g \psi G [ \overline{c} ]
\end{eqnarray}
This $G$ transformation also leaves the effective action invariant.
We then end up with a set of $F$ transformed BRST transformations
Eqs.(\ref{brst-f})
and a set of $G$ transformed anti-BRST transformations
Eqs.(\ref{brst-g}) and the abelian effective action.
It is clear that the $F$ transformed BRST and the $G$ transformed
anti-BRST
transformations are nilpotent. However the anticommutator of a $F$
transformed BRST
transformation Eqs.(\ref{brst-f}) with the $G$ transformed {\bf
anti}-BRST transformations Eqs.(\ref{brst-g}) does not vanish. If
we denote this anticommutator by $\Delta$ we obtain
\begin{eqnarray}
\label{delta}
&& \Delta A_\mu = i \partial_\mu (G F^t - F G^t)[ \lambda - \frac{1}{\xi}
\partial^\nu A_\nu ] \nonumber \\
&& \Delta c = \Delta \overline{c} = 0 \nonumber \\
&& \Delta \lambda = i \frac{1}{\xi} \Box ( G F^t - F G^t )[ \lambda -
\frac{1}{\xi} \partial^\mu A_\mu ] \nonumber \\
&& \Delta \psi = - i g \psi ( G F^t - F G^t ) [ \lambda - \frac{1}{\xi}
\partial^\mu A_\mu  ]
\end{eqnarray}
Notice that these transformations have ghost number zero. They do not
act on the ghosts and behave as gauge transformations on $A_\mu$ and
$\psi$.
We easily verify that  Eqs.(\ref{delta}) are a symmetry of the
effective action as well. We also find that they correspond to a  new
change of
variables defined by
\begin{eqnarray}
\label{h-transf}
&& A_\mu \rightarrow A_\mu + i \partial_\mu (H - H^t)[ \lambda -
\frac{1}{\xi} \partial^\nu A_\nu ] \nonumber \\
&& c \rightarrow c, \,\,\,\,\,\,  \overline{c} \rightarrow
\overline{c} \nonumber \\
&& \lambda \rightarrow  \lambda + \frac{i}{\xi} \Box ( H - H^t ) [
\lambda - \frac{1}{\xi} \partial^\nu A_\nu ]  \nonumber \\
&& \psi \rightarrow \psi - i g \psi ( H - H^t ) [\lambda - \frac{1}{\xi}
\partial^\nu A_\nu ]
\end{eqnarray}
Let us call this change of variables an $H$ transformation. They
correspond to the transformations Eqs.(\ref{delta}) with $H = G F^t -
F G^t$. This change of variables has Jacobian equals to one and also
leave the effective action invariant.

Then the freedom to perform changes of variables which leave
the ghost action and the path integral measure invariant reflects itself in
the BRST symmetry by allowing another set of transformations of the type
Eqs.(\ref{h-transf}). As a consequence the
anticommutator of the BRST and anti-BRST transformations does not need
to vanish and is proportional to an $H$ transformation Eqs.(\ref{h-transf}).

This can also be seen when we build the BRST charge in the BFV
formalism. In the abelian
case the BRST charge and the anti-BRST charge are
\begin{eqnarray}
\label{q}
Q &=& \int d^3 x \,\,\, ( \partial^i A_i \,\, c - i \Pi_0 {\cal P} )
\nonumber \\
\overline{Q} &=& \int d^3x \,\, (  \partial^i A_i \,\,\,\,
\overline{c} + i \Pi_0
\overline{{\cal P}} )
\end{eqnarray}
respectively. It is easily verified that they anticommute.
Now we can build the $F$ transformed BRST charge where the $F$
transformation on the ghost momenta is
defined as ${\cal P} \rightarrow F[ {\cal P} ] $ and $ \overline{\cal P}
\rightarrow ( F^{-1} )^t [ \overline{\cal P} ]$. We easily verify that
the $F$ transformation is a canonical transformation.
Now because the $F$ and $G$ transformations leave the effective action
invariant we could build the BRST charge with the $F$ transformed
ghosts and the anti-BRST charge with the $G$ transformed ghosts (the
$G$ transformation on
$\cal{P}$ and $ \overline{\cal P}$ being defined in a similar way to
the $F$ transformation). Then
the anticommutator of the BRST and anti-BRST charges no longer
vanishes and is proportional to $F G^t - F^t G$. We then see that in
the BFV formalism the changes in the ghosts variables we have been working
with correspond to canonical transformations on the ghosts. The
non vanishing of the
anticommutator of the BRST and anti-BRST transformations is due to
the fact that the charges are build with ghosts which have been
subject to different canonical transformations.

Since the effective action is invariant under BRST and anti-BRST
transformations it is also invariant under any linear combination of
them. If we take the original transformations the combined
transformation is nilpotent since each
of the original transformations is by itself nilpotent as is their
anticommutator. We could however consider the sum of the $F$
transformed BRST with
the $G$ transformed anti-BRST
transformations. We then have a set of
transformations which do not have a
well defined ghost number, leave the
effective action invariant and  are not  nilpotent because of
Eqs.(\ref{delta}).
This is the origin of the non-nilpotent symmetry found in
Ref.\cite{tang}.

\section{Conclusions}
\label{sec:end}

We have shown that the effective action of QCD in a covariant gauge
is invariant under non-local and even
not manifestly covariant BRST transformations either as a result of
the BFV path
integral formulation or as a  change of ghost variables which leave the
effective action and the path integral measure invariant. They just
reflect the
basic BRST symmetry in different forms. They are symmetries of
the full interacting quantum theory (in the absence of anomalies) but
do not entail any new
Ward identity besides those obtained from the usual BRST
transformations. It also shows the power of the Hamiltonian
formalism. Although the non standard form of the BRST transformations
can be found in the Lagrangian formalism its origin remains obscure
and its dependence on other known symmetries can not be traced. In the
BFV formalism all these issues can be clearly analysed.

It is worth remarking that it is possible for a gauge fixed action to
have further symmetries besides the BRST symmetry. One well known example
is the non-abelian Chern-Simons theory in Landau gauge in $2 + 1$
dimensions. It has a rigid vector supersymmetry \cite{birm} which is
independent of the BRST symmetry. In fact this vector
supersymmetry and the BRST symmetry
are part of a more general algebra which is a contraction of the
exceptional Lie superalgebra $D ( 2 | 1; \alpha )$ \cite{me}.
That this vector supersymmetry is a truly new symmetry manifests
itself in the existence of a new Ward identity which relates the
gauge and ghost inverse propagators \cite{birm}.

Our discussion on changes of ghost variables in Section \ref{sec:3}
were done only for the
abelian case and in the situation where the ghost transformations do
not
involve any field. In trying to consider a field dependent $F$
transformation we were led to very complicated expressions. Also the
non-abelian case became rather involved. We still miss a suitable
formalism to handle such situations.

\acknowledgments

The author wishes to thank P.van Nieuwenhuizen for conversations. This
work was partially supported by CNPq.


\begin{references}
\bibitem{conv} The space-time metric is $\eta_{\mu \nu} = diag (1, -1,
-1, -1)$. $\lambda^a$ are the generators of the Lie group. A dot means
time derivative and the brackets stand for Poisson
brackets.
\end{references}
\end{document}